\documentclass[pra,twocolumn,aps,amssymb,amsmath,superscriptaddress,showpacs]{revtex4-1}
\usepackage[latin1]{inputenc}
\usepackage{mathrsfs}
\usepackage{dsfont}
\usepackage{amsmath}
\usepackage{amssymb}
\usepackage{amsthm}
\usepackage{amsfonts}
\usepackage{amstext}
\usepackage{amsopn}
\usepackage{amsxtra}
\usepackage{mathrsfs}
\usepackage{dsfont}
\usepackage{color}
\usepackage{graphicx}

\newcommand{\ii}{\infty}
\newcommand\R{{\ensuremath {\mathbb R} }}
\newcommand\C{{\ensuremath {\mathbb C} }}

\newcommand\Z{{\ensuremath {\mathbb Z} }}

\newcommand\1{{\ensuremath {\mathds 1} }}
\newcommand\bS{{\ensuremath {\mathbb S} }}

\newcommand\nn{\nonumber}

\renewcommand\phi{\varphi}
\newcommand{\bH}{\mathbb{H}}

\newcommand{\cI}{\mathcal{I}}

\newcommand{\cE}{\mathcal{E}}

\newcommand{\cN}{\mathcal{N}}

\newcommand{\cH}{\mathcal{H}}

\newcommand{\eps}{\epsilon}

\newcommand{\bx}{\mathbf{x}}
\newcommand{\by}{\mathbf{y}}

\newcommand{\br}{\mathbf{r}}
\newcommand{\bp}{\mathbf{p}}
\newcommand{\bq}{\mathbf{q}}
\newcommand{\bk}{\mathbf{k}}

\renewcommand{\epsilon}{\varepsilon}
\newcommand\pscal[1]{{\ensuremath{\left\langle #1 \right\rangle}}}
\newcommand{\norm}[1]{ \left| \! \left| #1 \right| \! \right| }
\DeclareMathOperator{\tr}{{\rm Tr}}

\renewcommand{\geq}{\geqslant}
\renewcommand{\leq}{\leqslant}

\newcommand{\dx}{{\rm d}\bx}

\newcommand{\dy}{{\rm d}\by}

\newcommand{\dk}{{\rm d}\bk}

\newcommand{\kF}{k_{\rm F}}
\newcommand{\dpp}{{\rm d}\bp}
\newcommand{\dq}{{\rm d}\bq}

\date{\today}

\begin{document}

\title{Lower Bound on the Hartree-Fock Energy of the Electron Gas}

\author{David Gontier}
\affiliation{CEREMADE, Paris-Dauphine University, PSL University, 75016 Paris, France}

\author{Christian Hainzl}
\affiliation{Universit\"at T\"ubingen, Fachbereich Mathematik, Auf der Morgenstelle 10, 72 076 Tübingen, Germany}

\author{Mathieu Lewin}
\affiliation{CNRS \& CEREMADE, Paris-Dauphine University, PSL University, 75016 Paris, France}

\begin{abstract}
The Hartree-Fock ground state of the Homogeneous Electron Gas is never translation invariant, even at high densities. As proved by Overhauser, the (paramagnetic) free Fermi Gas is always unstable under the formation of spin or charge density waves. We give here the first explicit bound on the energy gain due to the breaking of translational symmetry. Our bound is exponentially small at high density, which justifies \emph{a posteriori} the use of the non-interacting Fermi Gas as a reference state in the large-density expansion of the correlation energy of the Homogeneous Electron Gas. We are also able to discuss the positive temperature phase diagram and prove that the Overhauser instability only occurs at temperatures which are exponentially small at high density. Our work sheds a new light on the Hartree-Fock phase diagram of the Homogeneous Electron Gas. 
\end{abstract}

\maketitle

The Homogeneous Electron Gas (HEG), where electrons are placed in a positively-charged uniform background, is a fundamental system in quantum physics and chemistry~\cite{ParYan-94,GiuVig-05}. In spite of its simplicity, it provides a good description of valence electrons in alcaline metals (\emph{e.g.} in solid sodium~\cite{Huotari-etal-10}) and of the deep interior of white dwarfs~\cite{Salpeter-61,BauHan-80}. It also plays a central role in the Local Density Approximation of Density Functional Theory~\cite{ParYan-94}, where it is used for deriving empirical functionals~\cite{Perdew-91,PerWan-92,Becke-93,PerBurErn-96}.

The ground state of the HEG is highly correlated at low and intermediate densities. It was first predicted by Wigner that the particles form a BCC ferromagnetic crystal at small densities~\cite{Wigner-34,Wigner-38}. 
But correlation also plays an important role at high densities:~The exact large-density expansion of the correlation energy has a peculiar logarithm due to the long range of the Coulomb potential, which cannot be obtained from regular second-order perturbation theory~\cite{Macke-50,BohPin-53c,Pines-53d,GelBru-57}.

In principle, the correlation energy of the HEG is defined as the difference between the Hartree-Fock (HF) ground state energy and the true energy. However, many authors use instead the (paramagnetic) non-interacting Fermi Gas as a reference. This state indeed provides the first two terms of the large-density expansion of the HEG total energy~\cite{GraSol-94}. But it is \emph{not} the absolute ground state of the Hartree-Fock HEG. This was first suggested by Wigner~\cite{Wigner-38} at high densities and then proved by Overhauser~\cite{Overhauser-60,Overhauser-62,Overhauser-68} who showed that the free Fermi Gas is unstable under the formation of spin or charge density waves. Recently, the phase diagram of the Hartree-Fock HEG has been studied numerically in great details~\cite{ZhaCep-08,BagDelBerHol-13,BagDelBerHol-14,Baguet-14}. It was discovered that the system is crystallized at all densities and that, at high densities, the electrons form an incommensurate lattice having more crystal sites than electrons~\cite{BagDelBerHol-13,Baguet-14}. Similar conclusions were reached in two space dimensions~\cite{BerDelDunHol-08,BerDelHolBag-11,Baguet-14}.

These works naturally raise the question of determining the energy gain of the true HF ground state, compared to the free Fermi Gas. A too large deviation could affect the large-density expansion of the exact correlation energy of the HEG. In~\cite[Eq.~(26)--(28)]{DelBerBagHol-15} it was argued that the Overhauser trial state only lowers the energy by an exponentially small amount:
\begin{equation}
e_{\rm HF}(r_s)- e_{\rm FG}(r_s) \lesssim -\frac{6.32\cdot 10^{-4}}{r_s^2}\exp\left(-\frac{23.14}{\sqrt{r_s}}\right). 
\label{eq:upper_bound}
\end{equation}
Here $e_{\rm HF}(r_s)$ is the exact (unknown) Hartree-Fock energy per particle and 
$$e_{\rm FG}(r_s)=\frac3{10}\left(\frac{9\pi}{4}\right)^{2/3}\frac1{r_s^2}-\frac3{4\pi}\left(\frac{9\pi}{4}\right)^{1/3}\frac{1}{r_s}$$ 
is the energy per particle of the (paramagnetic) free Fermi Gas. We work in terms of the dimensionless parameter $r_s=(3/4\pi\rho a_B^3)^{1/3}$ where $a_B=\hbar^2(me^2)$ is the Bohr radius. The energies are expressed in Hartree units, 1\;Ha$=\hbar^2/(ma_B^2)$. 

Based on the numerical simulations from~\cite{DelBerBagHol-15}, it seems plausible that the energy gain is indeed exponentially small. 
We provide here the first proof of this fact. More precisely, we show the exact inequality
\begin{multline}
e_{\rm HF}(r_s)-e_{\rm FG}(r_s)\\
\geq-\left(\frac{9\pi}{4}\right)^{\frac23}\frac{1+a\sqrt{r_s}}{r_s^2}\exp\left(-\frac{2^{-\frac13}3^{\frac13}\pi^{\frac23}}{\sqrt{r_s}}\right).
\label{eq:main_estimate} 
\end{multline}
The parameter $a$ can be adjusted as we like, but the bound~\eqref{eq:main_estimate} is only valid for $r_s$ smaller than a critical value $r_s(a)$, which tends to 0 when $a\to0$. Choosing for instance $a=4$, the condition is $r_s\leq 1.7$. Our estimate~\eqref{eq:main_estimate} takes exactly the same form as the upper bound~\eqref{eq:upper_bound} derived in~\cite{DelBerBagHol-15}, with however rather different constants. It confirms the prediction that the breaking of symmetry induces an exponentially small energy gain at large densities. In particular, our bound~\eqref{eq:main_estimate} gives the first justification of the use of the free Fermi Gas as a reference state in the large density expansion of the correlation energy of the HEG.  

Our proof of~\eqref{eq:main_estimate} proceeds in two steps. First, we bound the energy gain in terms of the lowest eigenvalue of an effective one-particle operator involving the Coulomb potential and a degenerate effective dispersion:
$$|P^2-\kF^2|-\frac{1}{r}.$$
Then, we estimate this eigenvalue using spectral techniques recently developed in the context of BCS theory~\cite{HaiSei-08b,FraHaiNabSei-07,HaiSei-16}. 

By slightly modifying the argument leading to the lower bound~\eqref{eq:main_estimate}, we are able to also estimate the critical temperature $T_c(r_s)$, above which the system is a paramagnetic fluid. We indeed show below that 
\begin{equation}
T_c(r_s)\leq \frac{5}4 \left(\frac{9\pi}4\right)^{\frac23}\frac1{r_s^2}\exp\left\{-\frac{2^{-\frac56}3^{\frac13}\pi^{\frac23}}{\sqrt{r_s}}\right\}
\label{eq:critical_temperature}
\end{equation}
for $r_s$ small enough. In addition, for the exponentially small temperatures where symmetry can be broken, we can prove that the gain in the free energy is also exponentially small, as it is for $T=0$.

The rest of the paper is devoted to the derivation of~\eqref{eq:main_estimate} and~\eqref{eq:critical_temperature}.

\section{A lower bound involving a degenerate Hydrogen-type operator}\label{sec:debut}

Let us consider a box $C_L$ of volume $L^3$, with periodic boundary conditions. We fix the Fermi level $\kF=(9\pi/4)^{1/3}r_s^{-1}$ and denote by $\gamma_{\rm FG}$ the corresponding free Fermi sea. Let $\gamma$ be the exact (unknown)  Hartree-Fock ground state in $C_L$, with the same number $N=2\#\{\bk\in (2\pi/L)\Z^3\ :\ k\leq\kF\}$ of electrons. The energy difference can be expressed as
\begin{multline}
\cE(\gamma)-\cE(\gamma_{\rm FG})=\sum_{\bk}\eps_L(\bk)\tr_{\C^2}\Big(\widehat{\gamma}(\bk,\bk)-\widehat{\gamma_{\rm FG}}(\bk,\bk)\Big)\\
-\frac{1}{2}\iint_{(C_L)^2}|\gamma(\bx,\by)-\gamma_{\rm FG}(\bx-\by)|^2G_L(\bx-\by)\,\dx\,\dy\\
+\frac{1}{2}\iint_{(C_L)^2}\rho_\gamma(\bx)\rho_\gamma(\by)G_L(\bx-\by)\,\dx\,\dy.
\label{eq:energy_difference}
\end{multline}
Here the $\C^2$--trace accounts for the spin summation and we use the notation $|A|^2:=\tr_{\C^2}(A^2)$ for a matrix $A$. The function $G_L$ is the $L$--periodic Coulomb potential with no zero mode, $\rho_\gamma(\bx)=\tr_{\C^2}\gamma(\bx,\bx)$ is the total density and $\widehat{\gamma_{\rm FG}}(\bk,\bk')_{\sigma,\sigma'}=\Theta(\kF-k)\delta(\bk-\bk')\delta(\sigma-\sigma')$. Finally,
$$\epsilon_L(\bk)=\frac{k^2}2-\frac{4\pi}{L^3}\sum_{\bp\neq0}\frac{\Theta(\kF-p)}{|\bp-\bk|^2}$$
is the $L$--periodic effective dispersion relation of the free Fermi Gas, which converges in the limit $L\to\ii$ to
\begin{equation*}
 \epsilon(k)=\frac{k^2}2-\frac{\kF}\pi \left(\frac{\kF^2-k^2}{2k\kF}\log\left|\frac{k+\kF}{k-\kF}\right|+1\right).
\end{equation*}
To arrive at the formula~\eqref{eq:energy_difference} we have expanded the exchange term and we have used that $\rho_{\gamma_{\rm FG}}$ is constant, hence the free FG energy has no direct term. 

We can replace $\eps_L$ by $\eps$ at the expense of an error $N\max_{\bk}|\eps_L(\bk)-\eps(k)|=o(N)$. Then we use the following expression for the first term in~\eqref{eq:energy_difference}
\begin{multline}
  \sum_{\bk}\eps(k)\tr_{\C^2}\Big(\widehat{\gamma}(\bk,\bk)-\widehat{\gamma_{\rm FG}}(\bk,\bk)\Big)\\
  = \sum_{\bk,\bk'}\left|\eps(k) -\eps(\kF)\right|\;\tr_{\C^2}\big|\widehat{\gamma}(\bk,\bk')-\widehat{\gamma_{\rm FG}}(\bk,\bk')\big|^2.
\label{eq:relative_kinetic_final_bound}
\end{multline}
Similar formulas have been used several times to control the exchange term in Hartree-Fock Quantum Electrodynamics~\cite{BacBarHelSie-99,HaiLewSerSol-07} and the variation of energy that an external potential can produce in a free Fermi sea~\cite{FraLewLieSei-11}. Formula~\eqref{eq:relative_kinetic_final_bound} follows from the remarks that $(i)$ for our two orthogonal projections we have 
$$(\gamma-\gamma_{\rm FG})^2=\gamma_{\rm FG}^\perp(\gamma-\gamma_{\rm FG})\gamma_{\rm FG}^\perp-\gamma_{\rm FG}(\gamma-\gamma_{\rm FG})\gamma_{\rm FG}$$ 
and $(ii)$ the dispersion relation is equal to $\pm|\eps(k)-\eps(\kF)|$ depending on whether $k\leq\kF$ or $k\geq\kF$ since $\eps$ is increasing. In other words, we use that 
$$\gamma_{\rm FG}=\Theta(\kF-k)=\Theta\big(\eps(\kF)-\eps(k)\big)$$
is also the ground state of its own effective dispersion $\eps$. 

It is useful to think of the relative density matrix 
$$\Psi(\bx,\by):=\gamma(\bx,\by)-\gamma_{\rm FG}(\bx,\by)$$ 
as a two-particle wavefunction with spin $1/2$ but without the fermionic or bosonic symmetry. Then, using~\eqref{eq:relative_kinetic_final_bound} we may rewrite the kinetic and exchange terms of~\eqref{eq:energy_difference} in the form $\pscal{\Psi|\bH_{2,L}|\Psi}$ with the two-particle operator
\begin{multline*}
\bH_{2,L}= \frac12\Big(\left|\eps(P_\bx) -\eps(\kF)\right|+\left|\eps(P_\by) -\eps(\kF)\right|\Big)\\ -\frac12 G_L(\bx-\by)
\end{multline*}
where $\mathbf{P}_\bx=-i\nabla_\bx$, hence $|\eps(P_\bx) -\eps(\kF)|$ is the operator in direct space corresponding to the degenerate dispersion relation $\bk\mapsto |\eps(k)-\eps(\kF)|$ for the Fourier coefficients.
The energy difference now reads 
\begin{multline}
\cE(\gamma)-\cE(\gamma_{\rm FG})=\pscal{\Psi\big|\bH_{2,L}\big|\Psi}\\
+\frac{1}{2}\iint_{(C_L)^2}\rho_\gamma(\bx)\rho_\gamma(\by)G_L(\bx-\by)\,\dx\,\dy+o(N).
\label{eq:energy_difference_ter}
\end{multline}
The two-particle Hamiltonian $\bH_{2,L}$ naturally describes the possible excitations of the free Fermi Gas (under the condition that the density is not altered to leading order, as it is the case for Spin Density Waves). It is the main object of interest for the Overhauser instability. In fact, $\bH_{2,L}$ is the Hessian of the Hartree-Fock energy at $\gamma_{\rm FG}$, to which the direct term has been dropped. The difficulty here is that we do not have the freedom to generate any two-particle wavefunction $\Psi(\bx,\by)$ that we like by perturbing $\gamma_{\rm FG}$. This is because $\Psi=\gamma-\gamma_{\rm FG}$ with $\gamma$ a one-particle density matrix, which implies some hidden constraints on $\Psi$. However, for a lower bound we may discard these constraints and simply bound 
$$\pscal{\Psi|\bH_{2,L}|\Psi}\geq \lambda_1(\bH_{2,L})\iint_{(C_L)^2}|\Psi|^2\geq 2N\lambda_1(\bH_{2,L}).$$
Here $\lambda_1(\bH_{2,L})$ is the (negative) ground state energy of $\bH_{2,L}$ and we have used that 
\begin{equation}
\iint_{(C_L)^2}|\Psi|^2=\sum_{\bk,\bk'}\tr_{\C^2}\Big|\widehat{\gamma}(\bk,\bk')-\widehat{\gamma_{\rm FG}}(\bk,\bk')\Big|^2\leq 2N.
\label{eq:normalization_Psi}
\end{equation}
After removing the center of mass we see that $\lambda_1(\bH_{2,L})=\lambda_1(\bH_{1,L})$ with the one-particle operator
$$\bH_{1,L}= \left|\eps(P_\br) -\eps(\kF)\right| -\frac12 G_L(\br).$$
Hence we have $\pscal{\Psi\big|\bH_{2,L}\big|\Psi}\geq 2N\lambda_1(\bH_{1,L})$.
For a lower bound we may discard the positive direct term in~\eqref{eq:energy_difference_ter} and, after passing to the thermodynamic limit $L\to\ii$, we arrive at our final lower bound on the relative energy per particle
\begin{equation}
e_{\rm HF}(r_s)-e_{\rm FG}(r_s)\geq 2\,\lambda_1\left(\left|\eps(P) -\eps(\kF)\right| -\frac1{2r}\right)
\label{eq:lower_bound_eigenvalue}
\end{equation}
with $P=-i\nabla_\br$.

\section{Study of the degenerate operator}
In this section we derive a bound on the lowest eigenvalue $\lambda_1(\bH_1)$ of the one-particle operator 
\begin{equation}
 \bH_1:=\left|\eps(P) -\eps(\kF)\right| -\frac1{2r},
 \label{eq:Hamiltonian_Hydrogen}
\end{equation}
appearing in~\eqref{eq:lower_bound_eigenvalue}. This is a Hydrogen-type Hamiltonian with the usual kinetic energy replaced by a dispersion relation degenerating on the Fermi sphere of radius $\kF$. 

First we replace $\eps$ by the non-interacting dispersion $k^2/2$ using that 
$$|\eps(k)-\eps(\kF)|\geq \frac12 |k^2-\kF^2|$$
for all $k$. This follows from the remark that if $f$ and $g$ are two increasing functions, then 
\begin{multline*}
|f(k)+g(k)-f(\kF)-g(\kF)|\\
=|f(k)-f(\kF)|+|g(k)-g(\kF)| \geq |f(k)-f(\kF)|.
\end{multline*}
This allows to remove the mean-field part in $\eps$, since it is increasing in $k$. After scaling we deduce that 
\begin{align}
\lambda_1\left(\bH_1\right)
\geq\frac{\kF^2}{2}\;\lambda_1\left(|P^2-1|-\frac{1}{\kF r}\right). \label{eq:simplify_H_1}
\end{align}
with of course $P^2=-\Delta_\br$. In the large--$\kF$ limit, we claim that the eigenvalue on the right behaves as
\begin{equation*}
\log\left[-\lambda_1\left(|P^2-1|-\frac{1}{\kF r}\right)\right]\underset{\kF\to\ii}\sim -C\sqrt{\kF}
\end{equation*}
for some $C$. Only the lower bound matters for our study. 

It is a general fact that operators of the form 
$$\left|P^2-1\right|-V(\br)$$ 
always have negative eigenvalues, whatever the size of the (attractive) potential~\cite{LapSafWei-02,HaiSei-10}. Operators of this type have appeared before in the study of the roton spectrum of liquid helium II~\cite{KruCol-01}
and in the BCS theory of superconductivity~\cite{HaiHamSeiSol-08,FreHaiSei-12,HaiSei-16,HaiLos-17}. Here we rely on techniques introduced in~\cite{FraHaiNabSei-07,HaiSei-08b,HaiSei-10} in the context of BCS theory. Since those in principle only apply to potentials decaying faster than the Coulomb potential at infinity, we first need to cut its long range, for instance  using the Yukawa potential. So we use the lower bound
\begin{equation}
 |P^2-1|-\frac{1}{\kF r}\geq |P^2-1|-\frac{Y_m(r)}{\kF}-\frac{m}{\kF}
 \label{eq:introduce_Yukawa}
\end{equation}
where $Y_m(r)=e^{-mr}/r$, and we estimate the first eigenvalue of $|P^2-1|-{Y_m(r)}/{\kF}$. The parameter $m$ will be chosen at the end of the argument.

In order to get our hands on the lowest eigenvalue, we use the Birman-Schwinger principle~\cite{LieSei-09}. It can be described as follows. Consider two positive operators $A,B\geq0$. Then 
$$(A-B)f=-Ef$$ 
if and only if 
$$C(E)g=g,\qquad\text{with}\quad C(E)=B^{\frac12}(A+E)^{-1}B^{\frac12}$$ 
and $g=B^{1/2}f$. Hence $-E$ is an eigenvalue of $A-B$ if and only if $1$ is an eigenvalue of $C(E)$. Now we remark that the eigenvalues of the operator $C(E)$ are decreasing with $E$. This implies that $E\geq -\lambda_1(A-B)$ if and only if all the eigenvalues of $C(E)$ are below 1, which is the same as saying that 
$$\norm{B^{1/2}(A+E)^{-1}B^{1/2}}_{\rm op}\leq 1$$ 
where $\|\cdot\|_{\rm op}$ denotes the operator norm (the largest eigenvalue). In our context, we deduce from this principle that
\begin{equation}
\norm{\sqrt{Y_m}\;\frac1{|P^2-1|+E}\;\sqrt{Y_m}}_{\rm op}\leq\kF
\label{eq:equality_norm}
\end{equation}
if and only if $E\geq -\lambda_1\big(|P^2-1|-\kF^{-1}Y_m(r)\big)$. This is how we are going to estimate the first eigenvalue from below. 

We now provide an upper bound on the operator norm in~\eqref{eq:equality_norm}. Following~\cite{FraHaiNabSei-07,HaiSei-08b,HaiSei-10}, we write the kernel of the operator in~\eqref{eq:equality_norm} in the form
$$\frac{1}{(2\pi)^3}\!\!\int_{0}^{\ii}\!\!\frac{r^2\,{\rm d}r}{|r^2-1|+E}\sqrt{Y_m(x)}\int_{\bS^2}e^{ir\omega\cdot(\bx-\by)}\,{\rm d}\omega\sqrt{Y_m(y)}$$
and estimate its norm by
\begin{multline*}
 \norm{\sqrt{Y_m}\;\frac1{|P^2-1|+E}\;\sqrt{Y_m}}_{\rm op}\\
 \leq \frac{1}{(2\pi)^3}\int_0^\ii\frac{r^2\cN(r)\,{\rm d}r}{|r^2-1|+E}
\end{multline*}
where 
\begin{align}
\cN(r)&:=\norm{\sqrt{Y_m(x)}\int_{\bS^2}e^{ir\omega\cdot(\bx-\by)}\,{\rm d}\omega\sqrt{Y_m(y)}}_{\rm op}\nn\\
&=\frac1{r^2}\norm{\sqrt{Y_{\frac{m}r}(x)}\int_{\bS^2}e^{i\omega\cdot(\bx-\by)}\,{\rm d}\omega\sqrt{Y_{\frac{m}r}(y)}}_{\rm op}.\label{eq:Nr}
\end{align}
The operator on the right acts by first multiplying by $\sqrt{Y_{m/r}}$, then going to the Fourier domain and restricting to the unit sphere, then going back to the direct space and multiplying again by $\sqrt{Y_{m/r}}$. The operator norm is the same as if we do things in the reverse order, namely we work with functions on the unit sphere that we multiply in space by $Y_{m/r}$. This uses the fact that the spectrum of $AA^\dagger$ is the same as that of $A^\dagger A$, except possibly for the eigenvalue $0$. From this we conclude as in~\cite{HaiSei-08b,HaiSei-10} that 
\begin{align*}
\cN(r)&=\frac{4\pi}{r^2}\max_{\int_{\bS^2}|f(\omega)|^2{\rm d}\omega=1}\;\int_{\bS^2}\int_{\bS^2}\frac{f(\bp)\,f(\bq)\,\dpp\;\dq}{|\bp-\bq|^2+(m/r)^2}\\
&=\frac{1}{r^2}\int_{\bS^2}\int_{\bS^2}\frac{\dpp\;\dq}{|\bp-\bq|^2+(m/r)^2}\\
&=\frac{4\pi^2}{r^2}\log\left(1+\frac{4r^2}{m^2}\right).
\end{align*}
In the second line we have used that the maximum is attained when $f$ is constant on the sphere, by~\cite[Rmk.~2.5]{HaiSei-10}. This is because $(\bp,\bq)\mapsto (|\bp-\bq|^2+(m/r)^ 2)^{-1}$ is rotationally invariant and pointwise positive, hence its highest eigenfunction can only be the trivial spherical harmonics, by the Perron-Frobenius theorem. 
As a conclusion, we have proved that 
\begin{multline*}
 \norm{\sqrt{Y_m}\;\frac1{|P^2-1|+E}\;\sqrt{Y_m}}_{\rm op}\\
 \leq\frac{1}{2\pi}\int_0^\ii\frac{{\rm d}r}{|r^2-1|+E}\log\left(1+\frac{4r^2}{m^2}\right):=\cI(E,m).
\end{multline*}
For small $E$ and small $m$, the integral behaves as
$$\cI(E,m)\underset{\substack{E\to0^+\\ m\to0^+}}{\sim}\frac1\pi \log(E^{-1})\log(m^{-1}).$$
More precisely, we have
\begin{equation}
\cI(E,m)\leq \frac1\pi \log(E^{-1})\log(m^{-1}) +C\log(m^{-1})+C
\label{eq:behavior_cI}
\end{equation}
for some large constant $C$ and for $E,m<1$. This behavior of $\cI$ suggests to take $E=e^{-\sqrt{\pi\kF}}$ and $m=t\sqrt{\kF}e^{-\sqrt{\pi\kF}}$ for some constant $t$. We then obtain 
\begin{equation*}
\cI\left(e^{-\sqrt{\pi\kF}},t\sqrt{\kF}e^{-\sqrt{\pi\kF}}\right)\leq \kF-\frac{\sqrt{\kF}}{2\sqrt{\pi}} \log\kF+O(\sqrt{\kF}).
\end{equation*}
Due to the logarithm, the right side is less than $\kF$ for $\kF$ large enough. We define $\kF(t)$ to be the smallest number for which $\cI(e^{-\sqrt{\pi\kF}},t\sqrt{\kF}e^{-\sqrt{\pi\kF}})\leq \kF$ for $\kF\geq\kF(t)$. Then the Birman-Schwinger principle gives 
\begin{equation*}
\lambda_1\left(|P^2-1|-\frac{Y_m(r)}{\kF}\right)\geq -e^{-\sqrt{\pi\kF}}
\end{equation*}
for all $\kF\geq\kF(t)$. Inserting in~\eqref{eq:introduce_Yukawa} we obtain
\begin{equation}
\lambda_1\left(|P^2-\kF^2|-\frac{1}{r}\right)\geq -\kF^2\left(1+\frac{t}{\sqrt{\kF}}\right)e^{-\sqrt{\pi\kF}}.
 \label{eq:bound_eigenvalue}
\end{equation}
Recalling~\eqref{eq:simplify_H_1} and~\eqref{eq:lower_bound_eigenvalue}, we obtain our final lower bound 
\begin{equation*}
e_{\rm HF}(r_s)-e_{\rm FG}(r_s)\geq -\kF^2\left(1+\frac{t}{\sqrt{\kF}}\right)e^{-\sqrt{\pi\kF}}. 
\end{equation*}
Inserting $r_s=(9\pi/4)^{1/3}\kF^{-1}$ and $a=t(9\pi/4)^{1/6}$, this is~\eqref{eq:main_estimate}.

In order to determine the concrete range of validity of our inequality, that is, the precise value of $\kF(t)$, we numerically solve the equation
$$\cI\left(e^{-\sqrt{\pi\kF(t)}},t\sqrt{\kF(t)}e^{-\sqrt{\pi\kF(t)}}\right)= \kF(t).$$
We have found for instance $r_s(2)\simeq 0.47$, $r_s(4)\simeq1.7$ and $r_s(10)\simeq5.5$.

\section{The critical temperature}\label{sec:critical_temperature}

We have seen that the energy gain due to the Overhauser instability is exponentially small at high density. Here we explain that our bound~\eqref{eq:bound_eigenvalue} on the effective one-particle operator $|P^2-\kF^2|-1/r$ can also be used to estimate the critical temperature. 

Let us go back to the box of side length $L$ and denote by $\gamma_{r_s,T,L}$ the paramagnetic Hartree-Fock fluid state of density $\rho$ at temperature $T>0$~\footnote{To be more precise, it is conjectured in~\cite{GonLew-18} that this state is unique for all $T,\rho>0$ and all $L$ large enough, but a rigorous proof is still missing. Our arguments here apply to any minimizer, in case there are several ones.}. This state is studied at length in~\cite{GonLew-18}. It solves the self-consistent equation
\begin{equation}
\widehat{\gamma_{r_s,T,L}}(\bk,\bk)=\bigg\{1+\exp\beta\Big(\eps_{r_s,T,L}(\bk)-\mu_{r_s,T,L}\Big)\bigg\}^{-1} 
\label{eq:SCF_box}
\end{equation}
with the dispersion relation
\begin{equation}
\eps_{r_s,T,L}(\bk)=\frac{k^2}2-\frac{4\pi}{L^3}\sum_{\bp\neq0}\frac{\widehat{\gamma_{r_s,T,L}}(\bp,\bp)}{|\bp-\bk|^2}
\end{equation}
and where the chemical potential $\mu_{r_s,T,L}$ is chosen to ensure that the total number of particles in the box is $N=\rho L^3$.
Then the free energy gain can be expressed similarly as in~\eqref{eq:energy_difference} in the form
\begin{align}
&\cE(\gamma)-TS(\gamma)-\cE(\gamma_{r_s,T,L})+TS(\gamma_{r_s,T,L})\nn\\
&=T\,\cH_{\rm FD}(\gamma,\gamma_{r_s,T,L}\big)\nn\\
&\ -\frac{1}{2}\iint_{(C_L)^2}\!|\gamma(\bx,\by)-\gamma_{r_s,T,L}(\bx-\by)|^2G_L(\bx-\by)\,\dx\,\dy\nn\\
&\ +\frac{1}{2}\iint_{(C_L)^2}\rho_\gamma(\bx)\rho_\gamma(\by)G_L(\bx-\by)\,\dx\,\dy
\label{eq:free_energy_difference}
\end{align}
where 
\begin{multline*}
\cH_{\rm FD}(A,B\big)\\
=\tr \Big\{A(\log A-\log B)+(1-A)(\log(1-A)-\log(1-B)) \Big\}
\end{multline*}
is the relative Fermi-Dirac entropy. In~\cite{HaiLewSei-08} (see also~\cite[Lemma~1]{FraHaiSeiSol-12}), it is proved that 
$$T\,\cH_{\rm FD}(A,B\big)\geq \tr\left\{ \frac{h}{\tanh(h/2T)}(A-B)^2\right\}$$
when $B=(1+e^{h/T})^{-1}$ is a Fermi-Dirac equilibrium state with one-particle Hamiltonian $h$. Using for instance
$$\frac{h}{\tanh(h/2T)}\geq T+\frac{|h|}{2}$$ 
and arguing as in Section~\ref{sec:debut}, we can control the energy gain from below by 
\begin{multline}
\cE(\gamma)-TS(\gamma)-\cE(\gamma_{r_s,T,L})+TS(\gamma_{r_s,T,L})\\
\geq \Big(T+\lambda_1\left(\mathbb{K}_{r_s,T,L}\right)\Big)\iint_{(C_L)^2}|\Psi|^2\label{eq:estim_positive_temp}
\end{multline}
with as before $\Psi(\bx,\by)=\gamma(\bx,\by)-\gamma_{r_s,T,L}(\bx,\by)$ and with, this time, the one-particle operator
$$\mathbb{K}_{r_s,T,L}=\frac12 \left|\eps_{r_s,T,L}(P_\br) -\mu_{r_s,T,L}\right| -\frac12 G_L(\br).$$
In the thermodynamic limit, this effective operator converges to
$$\mathbb{K}_{r_s,T}=\frac12 \left|\eps_{r_s,T}(P_\br) -\mu_{r_s,T}\right| -\frac1{2r}$$
with now $\eps_{r_s,T}(k)$ the self-consistent dispersion relation of the paramagnetic fluid state. In~\cite{GonLew-18} it is explained that $\eps_{r_s,T}(k)$ is radial increasing, as it was for $T=0$.

Our goal is to understand the region of the phase diagram where $r_s$ is small and 
\begin{equation}
T+\lambda_1\left(\mathbb{K}_{r_s,T}\right)>0.
\label{eq:implicit_condition_T}
\end{equation}
In this region we conclude from~\eqref{eq:estim_positive_temp} that the paramagnetic state $\gamma_{r_s,T,L}$ is the unique minimizer of the free energy, for $L$ large enough. Hence the free energies per particle satisfy
$$e_{\rm HF}(r_s,T)=e_{\rm HF,para}(r_s,T)$$
and the temperature $T$ is always above the critical temperature $T_c(r_s)$. Here $e_{\rm HF,para}(r_s,T)$ is the energy of the Hartree-Fock paramagnetic state, that is, the solution to the self-consistent equation~\eqref{eq:SCF_box} in the whole space $\R^3$ instead of the box $C_L$. We emphasize that  $e_{\rm HF,para}(r_s,T)$ differs from the Hartree-Fock energy of the free Fermi Gas at temperature $T$. On the other hand, in the region where $T+\lambda_1\left(\mathbb{K}_{r_s,T}\right)\leq0$, spin and charge density waves can form. We may however conclude, using~\eqref{eq:normalization_Psi}, that 
$$e_{\rm HF}(r_s,T)\geq e_{\rm HF,para}(r_s,T)-2\left|\lambda_1\left(\mathbb{K}_{r_s,T}\right)\right|.$$
Using similar arguments as for $T=0$, we will derive a bound on $\lambda_1\left(\mathbb{K}_{r_s,T}\right)$ which implies an estimate both on the free energy gain and on the critical temperature. 

The difficulty is that~\eqref{eq:implicit_condition_T} is an implicit condition linking $T$ and $r_s$. 
As a start we remark that the exchange term is maximized for the free Fermi gas at zero temperature, by rearrangement inequalities~\cite{LieLos-01}:
\begin{align}
\frac{1}{2\pi^2}\int_{\R^3}\frac{\widehat{\gamma_{r_s,T}}(\bp,\bp)}{|\bp-\bk|^2}\,\dpp&\leq \frac{1}{2\pi^2}\int_{\R^3}\frac{\widehat{\gamma_{r_s,T}}(\bp,\bp)}{p^2}\,\dpp\nn\\
&\leq\frac{1}{2\pi^2}\int_{\R^3}\frac{\Theta(\kF-p)}{p^2}\,\dpp\nn\\
&=\frac{2\kF}{\pi}=\frac{2^{\frac13}3^{\frac23}\pi^{-\frac23}}{r_s},\label{eq:estim_exchange}
\end{align}
an inequality which holds for all $\bk$. Next, in order to deal with all cases, we split the phase diagram into two regions, depending whether $\mu_{r_s,T}$ is smaller than $T$ or not.

Let us first consider the region where, for instance, $\mu_{r_s,T}\leq T$. 
Inserting this information in the self-consistent equation for the HF paramagnetic state $\gamma_{r_s,T}$, we find
\begin{align}
\rho= \frac{\kF^3}{3\pi^2}&=\frac1{(2\pi)^3}\int_{\R^3}\tr_{\C^2}\frac{1}{1+e^{\beta\left(\eps_{r_s,T}(k)-\mu_{r_s,T}\right)}}\,\dk\nn\\
&\leq \frac1{4\pi^3}\int_{\R^3} \frac{\dk}{1+e^{\beta\left(k^2/2-2\kF/\pi- T\right)}}\nn\\
&= \frac1{4\pi^3}\left(T+\frac{2\kF}{\pi}\right)^{\frac32}J\left(1+ \frac{2\beta\kF}{\pi}\right)\label{eq:estim_mu}
\end{align}
with 
$$J(\eta ):=4\pi\int_0^\ii \frac{r^2{\rm d}r}{1+e^{\eta (r^2/2-1)}}.$$
The function $J$ is decreasing and behaves as
$$\eta^{\frac32}J(\eta)\underset{\eta\to0}\sim 2(\sqrt2-1)\pi^{\frac32}\zeta\left(\frac32\right),\quad J(\eta)\underset{\eta\to\ii}\sim\frac{2^{\frac72}\pi}3.$$
Using the monotonicity of $J$ in~\eqref{eq:estim_mu} provides the bound
$$T\geq \left(\frac{4\pi}{3J(1)}\right)^{\frac23}\kF^2-\frac{2\kF}{\pi}.$$
With this information we can estimate $\lambda_1(\mathbb{K}_{r_s,T})$ by simply removing the absolute value. We obtain 
\begin{align}
T+\lambda_1\big(\mathbb{K}_{r_s,T}\big)&\geq T-\frac{\mu_{r_s,T}}2 +\frac12 \lambda_1\left(\eps_{r_s,T}(P_\br) -\frac1{r}\right)\nn\\
&\geq \frac{T}2 +\frac{\lambda_1(P^2/2-1/r)}{2}-\frac{\kF}{\pi}\nn\\
&\geq \frac12\left(\frac{4\pi}{3J(1)}\right)^{\frac23}\kF^2-\frac{2}{\pi}\kF-\frac14.\label{eq:final_1st_region}
\end{align}
This is positive for $\kF\gtrsim4.53$, that is, $r_s\lesssim0.42$. 
We can get a slightly better condition by evaluating the integral in~\eqref{eq:estim_mu} numerically. Namely, we first find the largest solution $\tau=\tau(\kF)$ to the implicit equation
$$\frac{\kF^3}{3\pi^2}= \frac1{4\pi^3}\left(\tau+\frac{2\kF}{\pi}\right)^{\frac32}J\left(1+ \frac{2\kF}{\tau\pi}\right)$$
and then ask when 
$$\frac{\tau(\kF)}2-\frac14-\frac{\kF}{\pi}>0$$
as required in~\eqref{eq:final_1st_region}. This provides the slightly better condition $\kF\gtrsim3.53$, or $r_s\lesssim0.54$, which we assume for the rest of the argument. Note that the condition can be further improved by taking $\mu\leq \eta T$ and optimizing over $\eta$ at the end, which we refrain from doing in order to keep our argument short. 

Next we turn to the region where the chemical potential satisfies $\mu_{r_s,T}\geq T>0$. Arguing as in~\eqref{eq:estim_mu} we find, this time,
\begin{align}
\rho= \frac{\kF^3}{3\pi^2}&\leq \frac{1}{4 \pi^3}\left(\mu_{r_s,T}+\frac{2\kF}{\pi}\right)^{\frac32}J\left(\beta\mu_{r_s,T}+ \frac{2\beta\kF}{\pi}\right)\nn\\
&\leq \frac1{4\pi^3}\left(\mu_{r_s,T}+\frac{2\kF}{\pi}\right)^{\frac32}J(1)\label{eq:better_J_infinity}
\end{align}
since $\beta\mu_{r_s,T}+ {2\beta\kF}/{\pi}\geq1$. This provides the lower estimate on the chemical potential
\begin{equation*}
\mu_{r_s,T}\geq\left(\frac{4\pi}{3J(1)}\right)^{\frac23}\kF^2-\frac{2\kF}{\pi}. 
\end{equation*}
We can get a similar upper bound by noticing that 
\begin{multline*}
\rho= \frac{\kF^3}{3\pi^2}\geq \frac1{4\pi^3}\int_{\R^3} \frac{\dk}{1+e^{\beta\left(k^2/2-\mu_{r_s,T}\right)}}\\
= \frac1{4\pi^3}\mu_{r_s,T}^{\frac32}J\left(\beta \mu_{r_s,T}\right)\geq \frac{1}{3\pi^2}\big(2\mu_{r_s,T}\big)^{\frac32}.
\end{multline*}
Altogether this proves that
\begin{equation}
\left(\frac{4\pi}{3J(1)}\right)^{\frac23}\kF^2-\frac{2\kF}{\pi}\leq \mu_{r_s,T}\leq \frac{\kF^2}{2}. 
\label{eq:mu_upper}
\end{equation}
In the second region, we have therefore shown that $\mu_{r_s,T}$ behaves essentially like $r_s^{-2}$, as it does for $T=0$. 

At this step we introduce the Fermi momentum $k_*$ such that $\eps_{r_s,T}(k_*)=\mu_{r_s,T}$. Note that, at $T=0$, $k_*$ is nothing else than $\kF$. We now demonstrate that $k_*$ behaves like $\kF$ in the region where $\mu_{r_s,T}\geq T$. 
We have
$$\mu_{r_s,T}=\frac{k_*^2}{2}-\frac{1}{2\pi^2}\int_{\R^3}\frac{\widehat{\gamma_{r_s,T}}(\bp,\bp)}{|\bp-\bk_*|^2}\,\dpp$$
so that, by~\eqref{eq:estim_exchange},
$$ \mu_{r_s,T}\leq\frac{k_*^2}{2}\leq \mu_{r_s,T}+\frac{2\kF}{\pi}.$$
Inserting~\eqref{eq:mu_upper}, this gives as we wanted
\begin{equation}
2\left(\frac{4\pi}{3J(1)}\right)^{\frac23}\kF^2-\frac{4\kF}{\pi}\leq k_*^2\leq  \kF^2+\frac{4\kF}{\pi}.
\label{eq:bound_k_star}
\end{equation}

The advantage of $k_*$ is that we can argue exactly as we did for $T=0$. Namely, in the absolute value 
$$|\eps_{r_s,T}(k)-\mu_{r_s,T}|=|\eps_{r_s,T}(k)-\eps_{r_s,T}(k_*)|$$ 
we may remove the monotone exchange term as for~\eqref{eq:simplify_H_1} and obtain
 \begin{align}
\lambda_1\left(\mathbb{K}_{r_s,T}\right)&\geq\frac{k_*^2}{4}\;\lambda_1\left(|P^2-1|-\frac{2}{k_* r}\right)\nn\\
&\geq-\frac{k_*^2}{4}\left(1+t\sqrt{\frac{2}{k_*}}\right)e^{-\sqrt{\pi k_*/2}}\label{eq:final_bd_eigenvalue_k_star}
\end{align}
where in the second line we have used our eigenvalue bound~\eqref{eq:bound_eigenvalue}. For $\kF\leq3.53$ we obtain from~\eqref{eq:bound_k_star} that $k_*\geq 1.76$, and we may for instance take $t=4(9\pi/4)^{1/6}$ (that is, $a=4$ in~\eqref{eq:main_estimate}). Then the function on the right side of~\eqref{eq:final_bd_eigenvalue_k_star} is decreasing and we may replace $k_*$ by its lower bound in~\eqref{eq:bound_k_star}. After a numerical evaluation of the multiplicative constant, we can conclude that the critical temperature is bounded above by
\begin{equation}
 T_c(r_s)\leq 0.68\;\kF^2\exp\left\{-\sqrt{\frac{c\pi\kF}2}\right\}
 \label{eq:first_bound_T_c}
\end{equation}
where 
$$c=\left(\frac{2^{\frac72}\pi}{3J(1)}\right)^{\frac13}<1.$$

With the first estimate~\eqref{eq:first_bound_T_c} on $T_c(r_s)$ we can get a better bound without the constant $c$, using the following argument. In the region where $T$ is less than the right side of~\eqref{eq:first_bound_T_c}, then $\beta\kF$ is exponentially large. Hence, going back to~\eqref{eq:better_J_infinity} we may replace $J(1)$ by $J(\ii)$, up to exponentially small errors. Then $J(1)$ gets also replaced by $J(\ii)$ in~\eqref{eq:final_bd_eigenvalue_k_star}, which replaces $c$ by 1 in the final estimate. To make this more quantitative, we may use for instance that in our region
\begin{align*}
J\left(\beta\mu_{r_s,T}+ \frac{2\beta\kF}{\pi}\right)&\geq J\left(\frac{2\beta\kF}{\pi}\right)\\
&\geq J(\ii)-7.48\,\kF\exp\left\{-\sqrt{\frac{c\pi\kF}2}\right\}
\end{align*}
since for instance $J(\eta)\geq J(\ii)-7/\eta$ for $\eta$ large enough. Replacing $J(1)$ in~\eqref{eq:bound_k_star} and using~\eqref{eq:final_bd_eigenvalue_k_star} together with a numerical evaluation of the multiplicative constant, we are now able to conclude, as we wanted, that 
\begin{align*}
T_c(r_s)&\leq \frac54 \kF^2\exp\left\{-\sqrt{\frac{\pi\kF}2}\right\}\\
&=\frac{5}4 \left(\frac{9\pi}4\right)^{\frac23}\frac1{r_s^2}\exp\left\{-\frac{2^{-\frac56}3^{\frac13}\pi^{\frac23}}{\sqrt{r_s}}\right\}.
\end{align*}
The numerical constant in the first line is $1.245$ which we have bounded by $5/4$ for simplicity. This concludes the derivation of our upper bound~\eqref{eq:critical_temperature} on the critical temperature $T_c(r_s)$. 

Let us finally consider the region of symmetry breaking which, for $r_s\lesssim0.54$,  is contained in the region where $\mu_{r_s,T}\geq T$ and $0\leq T\leq T_c(r_s)$. 
Our estimate on $\lambda_1\big(\mathbb{K}_{r_s,T}\big)$ in this region then provides immediately
\begin{multline*}
e_{\rm HF}(r_s,T)\geq e_{\rm HF,para}(r_s,T)\\-\frac{5}2 \left(\frac{9\pi}4\right)^{\frac23}\frac1{r_s^2}\exp\left\{-\frac{2^{-\frac56}3^{\frac13}\pi^{\frac23}}{\sqrt{r_s}}\right\}.
\end{multline*}
Since $T$ is exponentially small in this region, we may as well replace the (unknown) free energy $e_{\rm HF,para}(r_s,T)$ by the zero temperature energy $e_{\rm FG}(r_s)$ of the free Fermi Gas on the right side. This only generates another exponentially small error.

\section{Extension to the 2D Homogeneous Electron Gas}
Our argument is general and works similarly in 2D. The estimates~\eqref{eq:lower_bound_eigenvalue} and~\eqref{eq:simplify_H_1} are exactly the same. As in~\eqref{eq:introduce_Yukawa} we can bound $1/r\leq V_m(r)+Cm$ where, this time, $V_m$ is defined in Fourier space by $\widehat{V_m}(k)=(k^2+m^2)^{-1/2}$. The norm $\cN(r)$ in~\eqref{eq:Nr} is now given by an elliptic integral and the corresponding two-dimensional integral $\cI(E,m)$ satisfies the exact same asymptotics~\eqref{eq:behavior_cI}. Hence we find the same lower bound
\begin{equation*}
e_{\rm HF}^{\rm 2D}(r_s)-e^{\rm 2D}_{\rm FG}(r_s)
\geq-\kF^2\left(1+\frac{t}{\sqrt{\kF}}\right)e^{-\sqrt{\pi\kF}}.
\end{equation*}
An upper bound was first provided in 2D in~\cite[Eq. (44)]{BerDelDunHol-08} but a better bound can be derived following the arguments in~\cite{DelBerBagHol-15}. It is possible to estimate the critical temperature and the free energy gain by following the exact same argument as in Section~\ref{sec:critical_temperature}.

\section{Conclusion}

\begin{figure*}
\centering
\includegraphics[width=7cm]{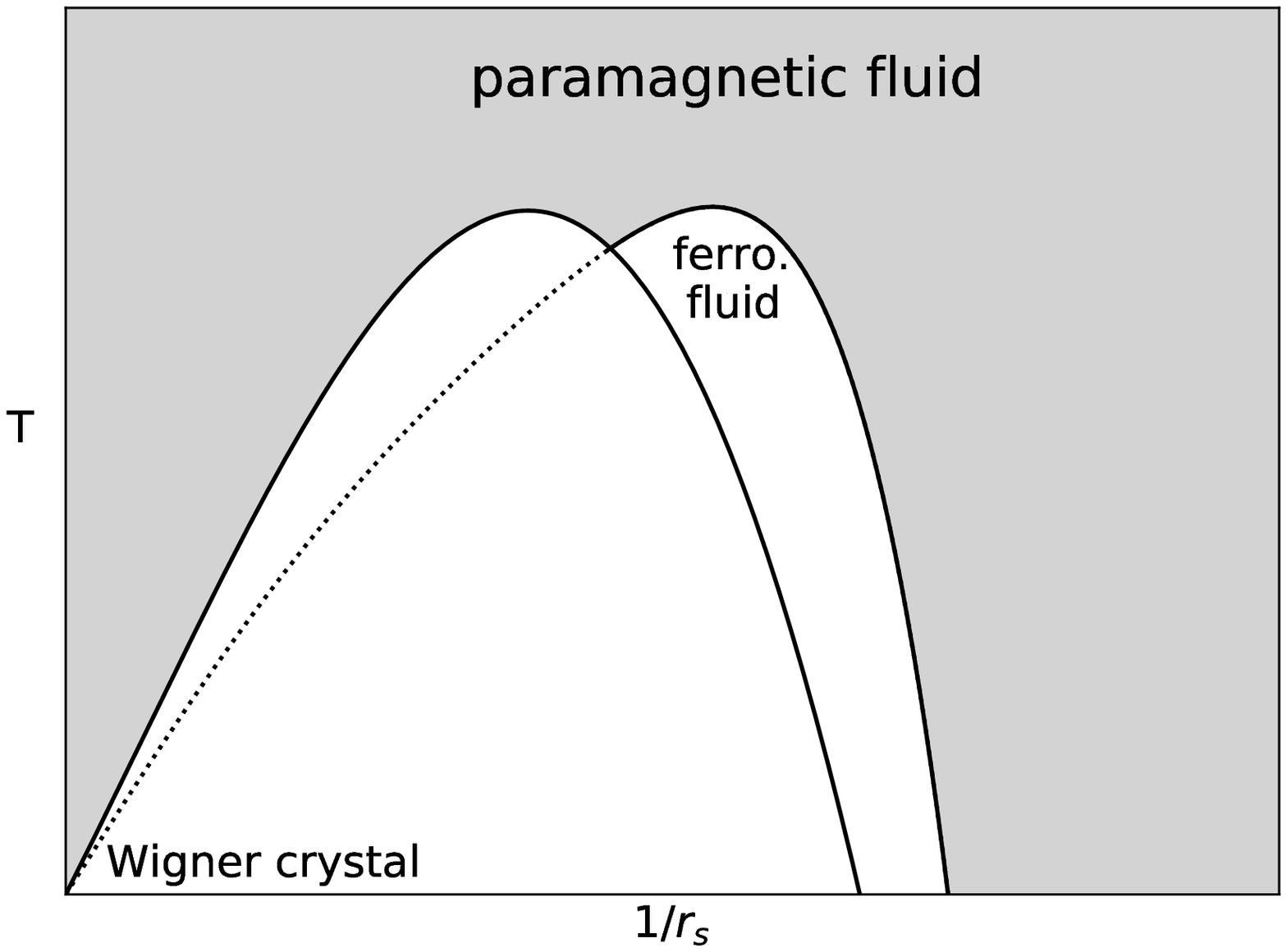}\hspace{1cm} \includegraphics[width=7cm]{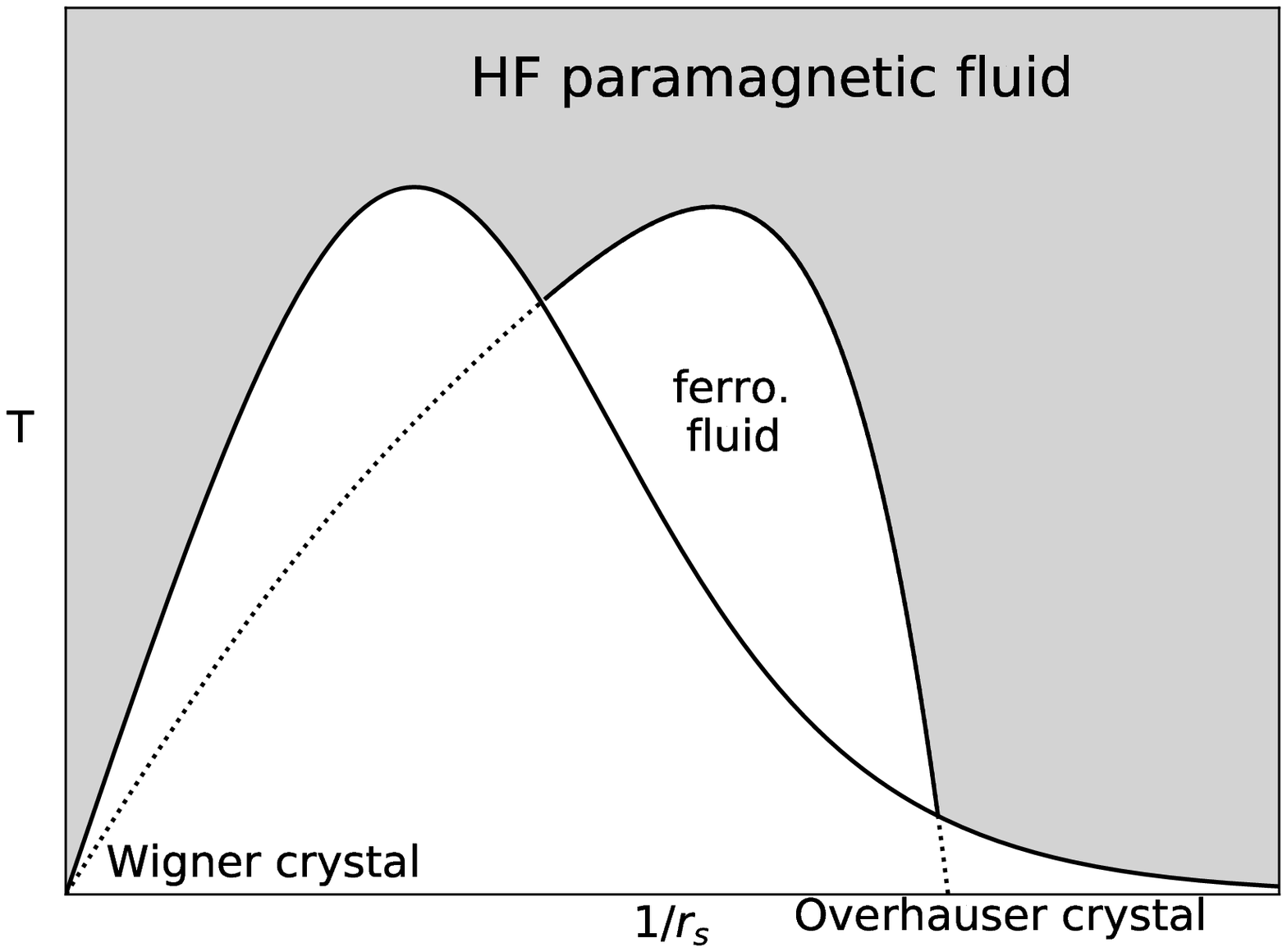}
 
\caption{\emph{Left:} General form of the true HEG phase diagram, as computed in~\cite{JonCep-96,ZonLinCep-02,DruRadTraTowNee-04,BroClaDubCep-13,FilForBonMol-15,SchGroVorBon-15,DorGroSjoMalFouBon-16,DorGroBon-18}. 
\emph{Right:} Expected general form of the Hartree-Fock HEG phase diagram, according to our work. The system is believed to be crystallized at all densities at $T=0$ but, as we prove in this paper, this ``Overhauser phase'' shrinks exponentially fast to the horizontal axis at large densities. The corresponding (free) energy gain is also exponentially small.\label{fig:phase_diagram}}
\end{figure*}

We have given the first rigorous proof that the breaking of translational symmetry in the Hartree-Fock Homogeneous Electron Gas can only decrease the Hartree-Fock ground state energy by an exponentially small amount at large density, as compared with the free Fermi Gas. In particular, the correlation energy can be defined by taking the FG as reference, up to an exponentially small error. In addition, we have also shown that the critical temperature (above which the gas is the paramagnetic fluid) is exponentially small at large densities. In the small region where symmetry breaking can happen, the free energy shift is also exponentially small.

An interesting question is to determine the precise asymptotics of the first eigenvalue of the degenerate Hydrogen-type Hamiltonian~\eqref{eq:Hamiltonian_Hydrogen}. It is however not clear if this eigenvalue can provide the exact behavior of the energy gain at $T=0$. Determining this gain in the large-density limit seems a very challenging task.

The Hartree-Fock phase diagram at $T=0$ was carefully computed in the recent works~\cite{ZhaCep-08,BagDelBerHol-13,BagDelBerHol-14,Baguet-14}. To our knowledge, much less is known about the full phase diagram at $T>0$. Our work gives the first indication that it has the general form displayed in Figure~\ref{fig:phase_diagram}. It is common wisdom that the Hartree-Fock model gives a very poor description of the HEG phase diagram. The ferromagnetic-to-paramagnetic fluid transition at $r_s\simeq 5.45$ and $T=0$ is sometimes mentioned as a major defect. But this transition actually does not exist, since the system is in a solid phase at this value of $r_s$. As is usual for nonlinear models, the use of more symmetry broken phases allows to slightly improve the energy. Although we have proved that this can only help by an exponentially small amount at very large densities, our estimates are too rough to conclude anything about what is happening at intermediate densities. This definitely calls for a more detailed numerical study of the HF phase diagram at positive temperature.

\bigskip

\noindent\textbf{Acknowledgments.} This project has received funding from the European Research Council (ERC) under the European Union's Horizon 2020 research and innovation programme (grant agreement MDFT No 725528 of M.L.). M.L. thanks Markus Holzmann for useful discussions.


%

\end{document}